\documentclass[prl,superscriptaddress,twocolumn]{revtex4-1}
\usepackage{tikz}
\usepackage{graphicx}
\usepackage{amsmath}
\usepackage{gensymb}
\usepackage[margin=1cm]{geometry}
\usepackage{setspace}
\usepackage{xcolor}
\usepackage{cancel}
\usepackage{soul}
\usepackage{wrapfig}
\usepackage{ulem}
\usepackage{units}
\usepackage{lipsum}

\newcommand{\TODO}[1]{\textcolor{black}{#1}}

\newcommand{\reftextit}[1]{}

\begin{document}

\title{Phonon-assisted inter-valley scattering determines ultrafast exciton dynamics\\
in MoSe$_2$ bilayers}

\author{Sophia Helmrich}
\affiliation{Department of Optics and Atomar Physics, Technical University Berlin, Berlin, Germany 10623}
\author{Kevin Sampson}
\affiliation{Department of Physics and Center for Complex Quantum Systems, The University of Texas at Austin, Austin, TX 78712, USA.}
\affiliation{Center for Dynamics and control of materials and Texas Materials Institute, 2501 Speedway, Austin, TX 78712, USA.}
\author{Di Huang}
\affiliation{Department of Physics and Center for Complex Quantum Systems, The University of Texas at Austin, Austin, TX 78712, USA.}
\author{Malte Selig}
\affiliation{Nichtlineare Optik und Quantenelektronik, Institut f\"ur Theoretische Physik, Technische Universit\"at Berlin, 10623 Berlin, Germany}
\author{Kai Hao}
\affiliation{Department of Physics and Center for Complex Quantum Systems, The University of Texas at Austin, Austin, TX 78712, USA.}
\author{Kha Tran}
\affiliation{Department of Physics and Center for Complex Quantum Systems, The University of Texas at Austin, Austin, TX 78712, USA.}
\author{Alexander Achstein}
\affiliation{Department of Optics and Atomar Physics, Technical University Berlin, Berlin, Germany 10623}
\author{Carter Young}
\affiliation{Department of Physics and Center for Complex Quantum Systems, The University of Texas at Austin, Austin, TX 78712, USA.}
\author{Andreas Knorr}
\affiliation{Nichtlineare Optik und Quantenelektronik, Institut f\"ur Theoretische Physik, Technische Universit\"at Berlin, 10623 Berlin, Germany}
\author{Ermin Malic}
\affiliation{Philipps University Marburg, Department of Physics, 35032 Marburg, Germany}
\author{Ulrike Woggon}
\affiliation{Department of Optics and Atomar Physics, Technical University Berlin, Berlin, Germany 10623}
\author{Nina Owschimikow}
\affiliation{Department of Optics and Atomar Physics, Technical University Berlin, Berlin, Germany 10623}
\author{Xiaoqin Li}
\affiliation{Department of Physics and Center for Complex Quantum Systems, The University of Texas at Austin, Austin, TX 78712, USA.}
\affiliation{Center for Dynamics and control of materials and Texas Materials Institute, 2501 Speedway, Austin, TX 78712, USA.}

\date{May, 2021}

\begin{abstract}

  While valleys (energy extrema) are present in all band structures of solids, their preeminent role in determining exciton resonances and dynamics in atomically thin transition metal dichalcogenides (TMDC) is unique. Using  two-dimensional  coherent  electronic  spectroscopy, we find that exciton decoherence occurs on a much faster time scale in MoSe$_2$ bilayers than that in the monolayers. We further identify two population relaxation channels in the bilayer, a coherent and an incoherent one. Our microscopic model reveals that phonon-emission processes facilitate scattering events from the $K$ valley to other lower energy $\Gamma$ and $\Lambda$ valleys in the bilayer. Our combined experimental and theoretical studies unequivocally establish different microscopic mechanisms that determine exciton quantum dynamics in TMDC monolayers and bilayers. Understanding exciton quantum dynamics provides critical guidance to manipulation of spin/valley degrees of freedom in TMDC bilayers. 
 
\end{abstract}
\maketitle

Much effort has been devoted to understanding the optical properties of semiconducting transition metal dichalcogenides (TMDCs) because of their unique layer-dependent band structures, strong light-matter interaction, and easy integration with other photonic structures \cite{2018_A.Chernikov&T.Heinz_ColloqiumRMP,2018_MullerMarlic_Review}. In both TMDC monolayers (MLs) and bilayers (BLs), exciton resonances dominate optical absorption spectra, exhibiting large oscillator strength and binding energy. These bright excitons correspond to direct transitions at the $K$ points and follow unique optical selection rules, often referred to as spin-valley locking \cite{2012_HL.Zeng&XD.Cui_NatNano_Valley,2012_K.F.Mak&F.Tony_NatNano_Valley,2012_D.Xiao&W.Yao_PRL_ValleyTheory,2012_TCao&GWang_NatComm_ValleyDichroism_MoS2}. The significantly stronger photoluminescence (PL) intensity in MLs compared to BLs indicates a transition from a direct to indirect band-gap \cite{2010_KF.Mak&F.Tony_PRL_MoS2,2010_A.Splendiani&F.Wang_NanoLett_MoS2}. Considering their similar absorption and markedly different PL, a question naturally arises: Is there any difference between the exciton quantum dynamics in TMDC MLs and BLs?

 Our study focuses on MoSe$_{2}$ MLs and BLs. The transition from a direct gap in ML MoSe$_{2}$ to an indirect gap in the BL coincides with the emergence of multiple low-energy valleys as illustrated in Fig.~1a. Valley scattering processes may strongly influence exciton quantum dynamics \cite{2019_MSelig&AKnorr_PRR_PhononValleyDarkExciton,2018_RA&CA_NanoLett_Ultrafast,2017_CChow&XXu_npj_phonon_exciton_dyn_MLMoSe2,2020_SBrem&EMalic_NanoLetters_PhononAssistPL}, which are characterized by two critical parameters: the population relaxation ($\Gamma=1/T_1$) and decoherence rates ($\gamma=1/T_2$). These quantum dissipative processes are related by $\gamma=1/T_{2}=1/2T_{1}+\gamma_{ph}$, where $\gamma_{ph}$ represents the pure dephasing.  Exciton quantum dynamics in TMDC monolayers have been investigated previously \cite{2015_MG&LXQ_NatComm_2DCS, 2016_HK&LXQ_NatPhys_2DCS, 2016_K.Hao&G.Moody_NanoLett_2DCS, 2018_M.Titze&HB.Li_PRMat_2DCS, 2018_Selig&Malic_2DMater, 2018_BremSelig&Malic_SciRep, 2020_Selig&Knorr_PRL}, and $K-K$ exciton coherence was found to be recombination-limited. In contrast, many questions related to exciton dynamics remain unknown in bilayers because of their more complex valley structure, layer pseudo-spins and indirect gaps \cite{2018_JLindlau&MSelig_NatComm_bilayerWSe2_valleys}. 

Here, we apply two-dimensional coherent electronic spectroscopy (2DCES) to investigate intrinsic exciton quantum dynamics in a MoSe$_2$ BL in comparison to a ML. Our measurements reveal ultrafast exciton decoherence time in MoSe$_2$ BLs to be $\sim$50 fs at low temperature, corresponding to a homogeneous linewidth of $2\gamma \sim\unit[27]{meV}$. This dephasing time is a factor of 6 shorter than that in the ML.  We further distinguish two distinct population relaxation channels, a coherent population relaxation occurring on a time scale of $\sim$55 fs, and an incoherent population relaxation that occurs on $\sim$800 fs.  Microscopic calculations yield excellent agreement with experiments and suggest that the ultrafast exciton dephasing and population relaxation in the BL originate from  phonon-assisted inter-valley scattering processes from the $K$ valley to other, lower energy valleys (i.e. $\Lambda$ and $\Gamma$ valleys). An enhanced exciton decoherence arising from inter-valley scattering is likely also present in other stacked and twisted TMDC BLs with momentum-indirect states below the optically-bright excitons \cite{2020_Park_Broken_Mirror,2021_Lukin_reconstructed_moire}.

The MoSe$_2$ ML and BL are mechanically exfoliated from a bulk crystal and transferred to a sapphire substrate for optical measurements (more details in SI). All optical measurements are performed at $\sim$\unit[30]{K} unless otherwise specified. We observe two resonances in both the ML and BL in linear reflectivity measurements and attribute them to the A and B excitons. The A exciton is red-shifted in the BL but the B resonance energy is nearly constant. \TODO{This observation is consistent with earlier experiments and confirms the spectral uniformity of sample~\cite{2015_A.Arora&M.Potemski_Nanoscale_MoSe2, niu2018thickness}.} We extract a full width half maximum (FWHM) of $\sim$\unit[45]{meV} and $\sim$\unit[80]{meV} for the ML and BL, respectively, by fitting with a Voigt function. The dominant contribution to the exciton linewidth  at low temperature in linear spectroscopy is inhomogeneous broadening.

\TODO{The lowest two conduction and valence bands calculated from density functional theory (DFT) are displayed in Fig.~1a. The A exciton corresponds to the $K-K$ transition between the first valence band and the lowest conduction band while the energy splitting between the A and B excitons mostly results from the strong spin-orbit interaction in TMDCs~\cite{2014_Roldan}. Critically, these and other~\cite{2020_Pandey} DFT calculations show the emergence of lower-energy valleys in BLs, which leads to increased intervalley scattering and dramatically alters exciton quantum dynamics, as we show below. Although the energy of other valleys relative to the $K$ points is important to our theoretical model, the absolute transition energy in DFT calculations cannot be directly compared to experimentally observed exciton resonances because DFT routinely underestimates band gaps~\cite{perdew2017understanding} and exciton binding energies are not included.}  

\begin{figure}
\centering
\includegraphics[width=0.48\textwidth]{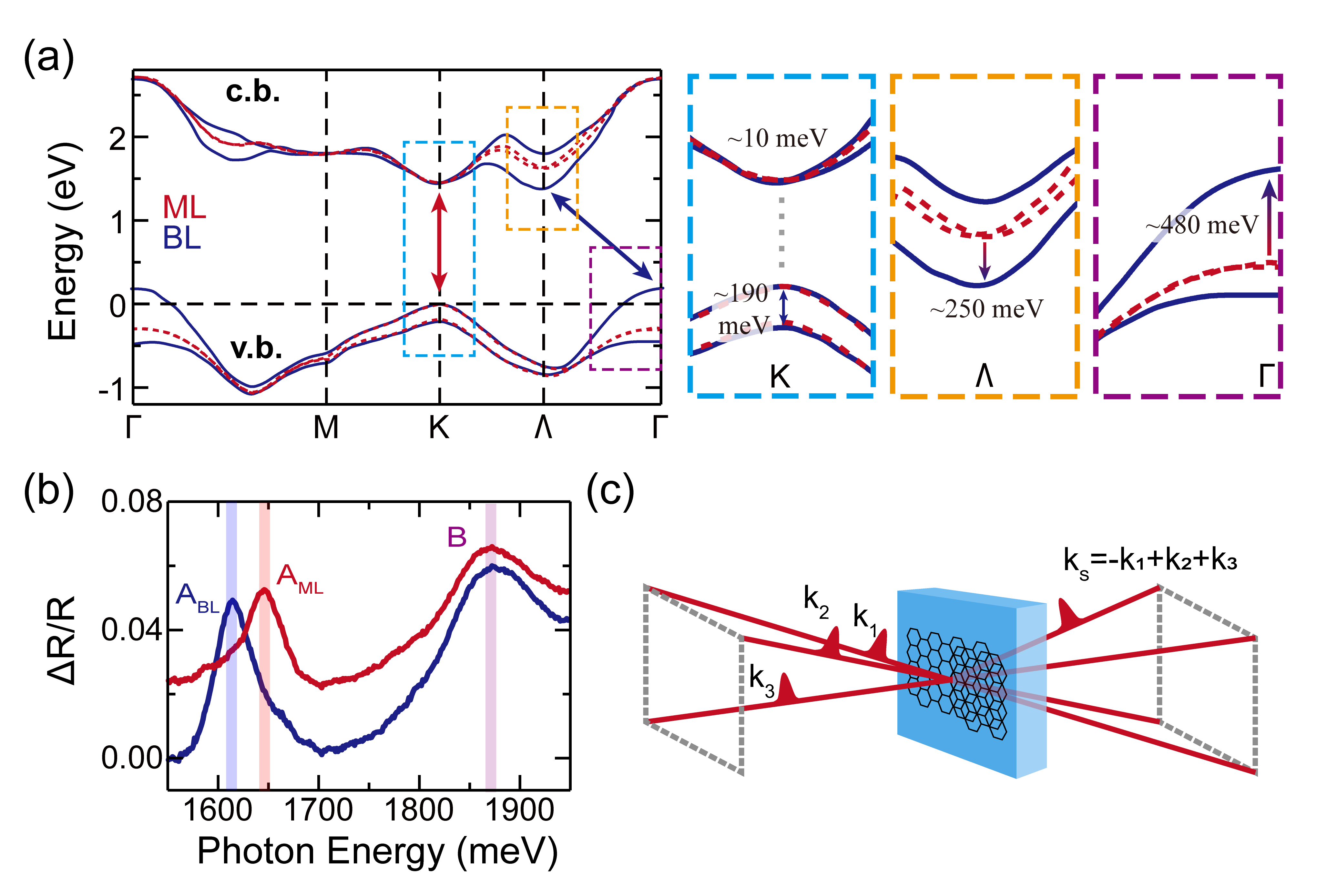}
\caption{\label{Fig1} (a) Single-particle band structures of MoSe$_2$ ML (red dashed lines) and BL (blue solid lines) with spin-orbit coupling, showing the two highest valence and the lowest conduction bands adapted from Roldan et al. \cite{2014_Roldan}. Details of the band evolution from ML to BL at the $K,\Lambda$ and $\Gamma$ points are shown in the blue, orange and purple rectangles at right. (b) Reflectance spectra for MoSe$_2$ ML (red) and BL (blue) at \unit[30]{K}. (c) Schematic of the 2DCES experiment in a box geometry.}
\end{figure}

The 2DCES experimental set-up has been described in detail in previous studies \cite{2015_MG&LXQ_NatComm_2DCS, 2016_HK&LXQ_NatPhys_2DCS, 2017_K.Hao&G.Moody_NatComm_2DCS,2018_M.Titze&HB.Li_PRMat_2DCS}. Briefly, three phase-stabilized, co-circularly polarized excitation laser pulses are derived from the same Ti:Sapphire laser with $\sim$\unit[60]{fs} pulse duration and \unit[76]{MHz} repetition rate, with adjustable time delays ($t_1$ and $t_2$) between them. We choose the co-circular polarization for all pulses to resonantly excite excitons in one $K$ valley. The three beams are arranged in the standard box-geometry shown in Fig.~1c and focused to a single spot $\sim$\unit[8]{$\mu$m} in diameter. The photon-echo or four-wave mixing signal is generated along the fourth corner of the box, characterized by wavevector $k_S= -k_1+k_2+k_3$. Both the amplitude and phase of the nonlinear signal are measured via spectral interference with a fourth reference pulse separated by a time delay $t_3$ from the third pulse.

We first investigate exciton decoherence by taking the one-quantum rephasing spectrum. As shown in Fig.~2a, the one-quantum rephasing spectrum is obtained by scanning $t_1$ and $t_3$ while keeping $t_2$ constant. The time-domain signal is converted to the frequency domain via Fourier transform. For the measurement presented here, $t_2=\unit[0]{fs}$ is chosen. Elongation along the diagonal of the 2D spectrum is due to inhomogeneous broadening from variations in strain and dielectric environment, impurities, or defects \cite{2019_TJakubczyk&JKasprzak_InhomogeneousBroadening}. In contrast, the cross-diagonal broadening along $\hbar\omega_{t_1} = -\hbar \omega_{t_3}$ reveals the intrinsic homogeneous linewidth $\gamma$, which is inversely proportional to the dephasing time $1/T_2=\hbar\gamma$~\cite{2010_SiemensHeinz_OpticExpress}.

The monolayer spectrum Fig.~2b features two prominent diagonal peaks attributed to the neutral exciton X$^0$ (\unit[1652]{meV}) and trion X$^T$ (\unit[1625]{meV}) \cite{2013_JS.Ross&XD.Xu_NatComm_Trion, 2016_K.Hao&G.Moody_NanoLett_2DCS} and coherent coupling between excitons and trions can be identified through the cross peaks. The A exciton homogeneous broadening $\gamma_{ML}^X=\unit[2.1\pm0.2]{meV}$ ($T_{2,ML}^X =\unit[313\pm33]{fs}$) is extracted from a Lorentzian fit for the exciton peak, shown in Fig.~2c. All these features are consistent with previous studies \cite{2016_K.Hao&G.Moody_NanoLett_2DCS, 2018_M.Titze&HB.Li_PRMat_2DCS,2008_MCho_Giant_2D_Review_Chem}.

\begin{figure}
\centering
\includegraphics[width=0.48\textwidth]{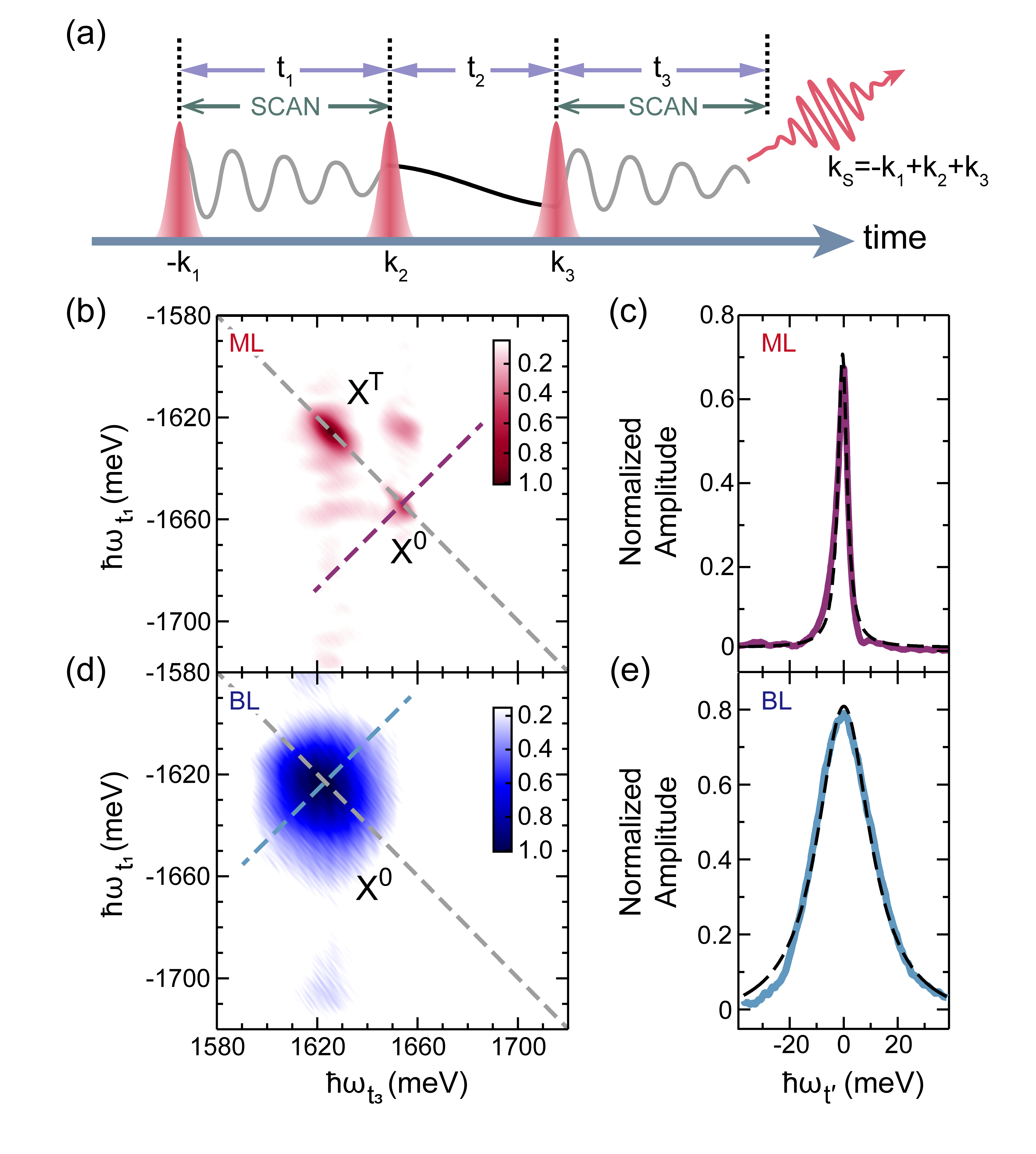}
\caption{\label{Fig2} {One-quantum rephasing spectra from MoSe$_2$ ML and BL. (a) Schematic showing the one-quantum rephasing pulse sequence. (b, d) Amplitude spectra of a MoSe$_2$ ML (BL) at $\unit[1\times10^{12}]{cm^{-2}}$ excitation density and \unit[30]{K}. The exciton and trion resonances are indicated by X$^0$ and X$^T$ in the ML. The cross-diagonal linewidth (homogenous linewidth) is extracted at the X$^0$ peak indicated by the dotted line. (c, e) The extracted homogeneous linewidths are fitted with Lorentzian functions for ML and BL MoSe$_2$ respectively. Here, $\omega_{t'}=\omega_{t_1}+\omega_{t_3}$}.}
\end{figure}

In the 2D spectrum taken from the BL in Fig.~2d, only one diagonal peak corresponding to the A exciton at \unit[1625]{meV} is observed over the spectral range covered by the excitation laser pulse (\unit[$1635\pm20$]{meV}). In contrast to the ML, the line shape of the exciton resonance in the MoSe$_2$ BL is nearly homogeneously broadened. Following a similar analysis, we extract a homogeneous broadening in Fig.~2e of $\gamma^X_{BL}= \unit[13.6\pm0.8]{meV}$ ($T_{2,BL}^X=\unit[49\pm2]{fs}$). The excitons in bilayer MoSe$_2$ exhibit $\sim$ 6 times faster dephasing than the ML A excitons. While different substrates (e.g hBN) can alter exciton dephasing in ML by suppressing charge fluctuations or modifying the photon density~\cite{martin2020encapsulation}, we anticipate a reduced substrate influence on BLs because of the rapid inter-valley scattering.

\begin{figure}
\centering
\includegraphics[width=0.48\textwidth]{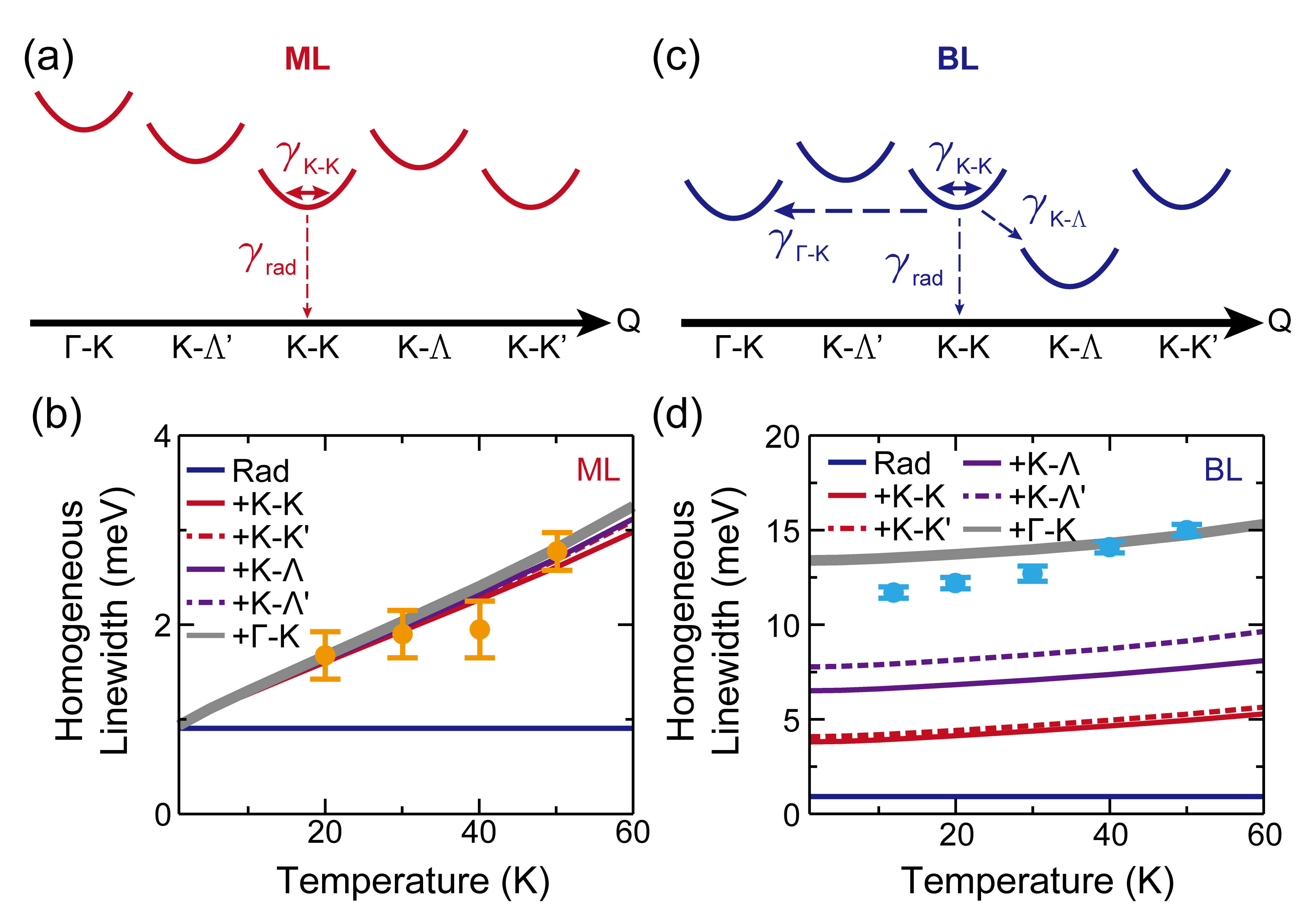}
\caption{\label{Fig3} Exciton dephasing as a function of temperature. (a, c) Illustration of valley scattering processes influencing exciton dephasing in MoSe$_2$ (a) MLs and (c) BLs. The horizontal axis Q stands for center-of-mass momentum, and vertical axis represents exciton energy. (b,d) Calculated dephasing channels for ML (b) and BL (d), respectively. Linewidth broadening due to contributions from radiative decoherence, exciton intravalley scattering ($K-K$) and intervalley scattering to $K-K'$, $K-\Lambda$, $K-\Lambda’$ and $\Gamma-K$ are accumulated in each curve stacked vertically. Experimental homogeneous linewidths (extrapolated to zero-excitation-density) are shown as tangerine (ML) and blue (BL) points.}
\end{figure}

To reveal the origin of exciton decoherence, we apply a microscopic theory that quantitatively evaluate the role of exciton-phonon interaction and inter-valley scattering \cite{ 2016_M.Selig&A.Knorr_NatComm_CoherenceTheory,2018_RA&CA_NanoLett_Ultrafast}. We start with the different ML and BL electronic band structures (Fig.~1a) from first-principle calculations, supported by ARPES experiments.~\cite{ 2014_Y.Zhang&ZX.Shen_NatNano_ARPES, 2014_JG&Cesare_PRB_theory} We then include excitonic effects by solving the Wannier equation, taking the modified Coulomb potential in ML and BL MoSe$_2$ into account  \cite{2016_M.Selig&A.Knorr_NatComm_CoherenceTheory,2019_OvsenMalic_CommunPhys}. The lowest lying exciton state $\nu = 1s$ is described by 
\begin{equation}
E^{\xi_h\xi_e}_\mathbf{Q} = E^{\xi_h\xi_e}_0 + E^{\xi_h\xi_e\,1s}_B + \frac{\hbar^2\mathbf{Q}^2}{2M^{\xi_h\xi_e}},
\end{equation}
where the first term accounts for the energetic separation of the different intra- and intervalley transitions in the electronic picture, the second term accounts for the binding energy of the respective transition and the third term accounts for the kinetic energy of the exciton with an effective mass $M^{\xi_h\xi_e} = m_h^{\xi_h} + m_e^{\xi_e}$. 

The low-energy excitons consist of electrons and holes located at several high-symmetry points $\xi_{e/h}$ in the Brillouin zone, namely $K$($K'$), $\Lambda$($\Lambda'$) points for electrons and $K$($K'$), $\Gamma$ points for holes, respectively. The three-fold rotational symmetry leads to energetically degenerate $K'$ and $\Lambda'$, points with opposite spins. We summarize the relevant exciton states in Fig.~3a (ML), 3c (BL) and Table.~1 in SI. In ML MoSe$_2$, the lowest-energy exciton transition is a direct transition at the $K-K$ point. In contrast, the band structure of BL MoSe$_2$ evolves from a direct to indirect band gap, with the valance band maximum (VBM) shifting from the $K$($K'$) point to the $\Gamma$ point and the conduction band minimum (CBM) shifting from the $K$($K'$) point to the $\Lambda$($\Lambda'$) point. The drastic band structure evolution from monolayer to bilayer is attributed to the fact that conduction band at the $\Lambda$($\Lambda'$) point and valence band at the $\Gamma$ point are primarily composed of out-of-plane orbitals, while bands at the $K$ point are mainly composed of in-plane orbitals \cite{2020_Pandey}. The valley indirect $\Gamma-K$, $K-\Lambda$ and $\Gamma$-$\Lambda$ excitons are unobservable in the reflectivity and 2D spectra because of their significantly reduced oscillator strength. \TODO{The key difference between the ML and BL is the emergence of the low-energy valleys in BL. Energetically favorable  valley scattering processes become the dominant channel of A exciton decoherence in the BL even at low temperature and lead to $\sim$ 6 times faster dephasing than that found in MLs.}  

We quantitatively evaluate several decoherence channels of the bright $K-K$ excitons. By solving Maxwell and Bloch equations, and performing a correlation expansion for the exciton-phonon interaction in the second-order Born-Markov approximation \cite{2018_Selig&Malic_2DMater}, we first calculate the radiative decay process described in a previous study \cite{ 2016_M.Selig&A.Knorr_NatComm_CoherenceTheory}. Here, we focus on the phonon-assisted decoherence rate in ML and BL \cite{ 2016_M.Selig&A.Knorr_NatComm_CoherenceTheory, 2018_RA&CA_NanoLett_Ultrafast}:
\begin{align}
\gamma^{K-K}_{phon} &= \sum_{\mathbf{Q},i,\alpha,\pm}|g_\mathbf{Q}^{K-K\rightarrow i}|^2 \left( \frac{1}{2} \pm \frac{1}{2} + n^{K-i\,\alpha}_\mathbf{Q} \right) \times \nonumber\\ &\times \mathcal{L}_{\gamma} \left( E_\mathbf{Q}^{i}  - E_\mathbf{0}^{K-K} \pm \hbar \Omega^{K-i\,\alpha}_\mathbf{Q} \right). \label{gamma_phon}
\end{align}
The summation $i$ incorporates all possible excitonic valleys. In particular it incorporates intravalley scattering ($i=K-K$), intervalley scattering via electron scattering ($i=K-K',K-\Lambda,K-\Lambda'$) and intervalley scattering via hole scattering ($i=K'-K$). 
The $\pm$ sum accounts for phonon emission and absorption processes, $n^{\xi\,\alpha}_\mathbf{Q}$ and $\hbar \Omega^{\xi\,\alpha}_\mathbf{Q}$ account for the phonon occupation and the phonon dispersion at the $\xi$ point in the Brillouin zone and branch $\alpha$ \cite{2014_ZH.Jin&KW.Kim_PRB}. In the calculation, we include the LA, TA, LO, TO and A$'$ modes which provide the strongest coupling strength in monolayer TMDCs \cite{2014_ZH.Jin&KW.Kim_PRB}. 
In this study focusing on quantum decoherence effects in BLs, we assume that the exciton-phonon coupling elements $g_\mathbf{Q}^{K-K\rightarrow i\,\alpha}$ appearing in Eq.~\ref{gamma_phon} can be approximated with the according values for the ML material (see SI). The Lorentzian $\mathcal{L}_{\gamma}(\Delta E)$ with broadening $\gamma$ accounts for the relaxed energy conservation during an exciton-phonon scattering event, while the broadening $\gamma$ is  calculated by self-consistently solving equation (2) \cite{martin2020encapsulation}.

The results of the calculation are summarized in Fig.~3b where each curve plots the accumulative contribution to the linewidth. As an example, the red curve labeled $+K-K$ is a sum of the contribution from radiative decay and the intravalley exciton scattering within the $K$ valleys (see SI for more details). In ML MoSe$_2$, the dephasing rate is mainly determined by the radiative decay and intravalley phonon scattering \cite{2016_TJakubczyk&JKasprzak_NanoLetters_FWMMicroscopy}. At low temperatures, the dephasing rate increases linearly with temperature due to the absorption and emission of long range acoustic phonons \cite{2016_PDey&DKaraiskaj_PRL_Optical_Coherence_Phonon}. The contribution from intra-valley phonon induced decoherence approaches zero as temperature approaches zero. In contrast, both calculated and measured homogeneous linewidths in BL MoSe$_2$ remain broad $\sim$ \unit[14]{meV} even in the low temperature limit, as shown in Fig.~3d. The self-consistent solution of Eq.~\ref{gamma_phon} reveals that the dominant process is exciton scattering from $K-K$ exciton to $\Gamma-K$ states via emission of acoustic and optical $K$ phonons. Such a phonon emission process remains efficient even at low temperatures. The coupling to optical phonons here in MoSe$_2$ is stronger than that in WS$_2$, previously studied using linear spectroscopy methods \cite{2018_RA&CA_NanoLett_Ultrafast}. After investigating exciton-exciton interactions via excitation power dependent measurements (details included in SI), we directly compare the extrapolated and calculated homogeneous linewidth (Fig.~3b and 3d) in the relevant temperature range, finding remarkable agreement within $\sim$20\%. This agreement suggests that the calculations have captured the most important quantum decoherence mechanisms in both the ML and BL.

\begin{figure}
\centering
\includegraphics[width=0.48\textwidth]{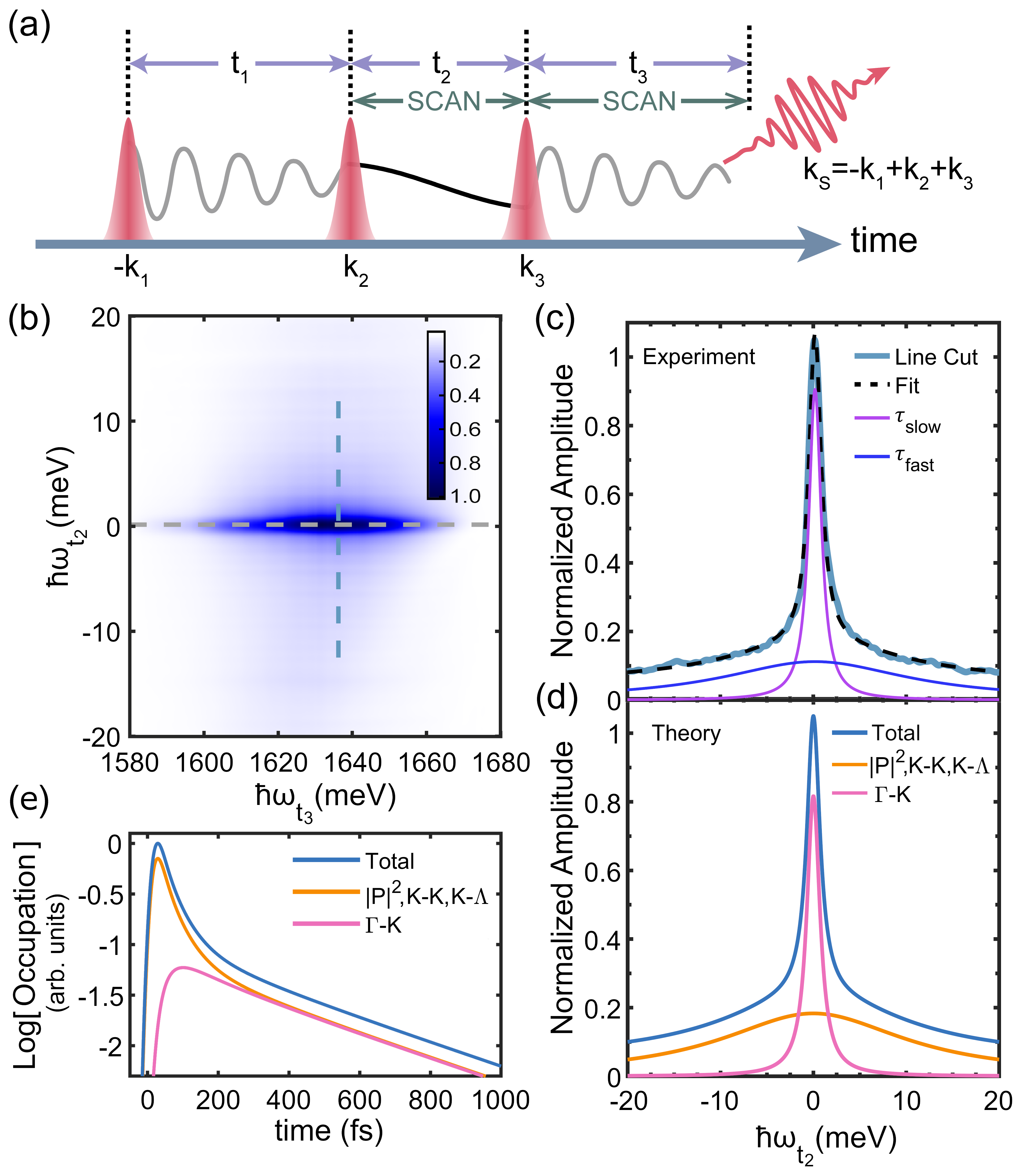}
\caption{\label{Fig4} Zero-quantum spectra from a MoSe$_2$ BL used to extract population relaxation times. (a) Schematic showing the zero-quantum pulse sequence. (b) Zero-quantum spectrum of a MoSe$_2$ BL at $\unit[1\times10^{12}]{cm^{-2}}$ excitation density and \unit[30]{K}. The vertical line cut at the peak of the X$^0$ resonance captures population relaxation dynamics. (c) Fitting the zero-quantum line cut with two Lorentzian functions reveals fast $\unit[54\pm2]{fs}$ and slow $\unit[810\pm10]{fs}$  decay components. (d) Calculated relaxation dynamics with two dominant components in the frequency domain in excellent agreement with experiment. (e) Theoretical calculation of time-domain population dynamics in the $K$ valley after excitation of $K-K$ excitons. In (c-e), totals are offset from the components for clarity.}
\end{figure}

Next, we extract the exciton population relaxation dynamics by taking zero-quantum spectra. These spectra S($\rm{t_1},\hbar\omega_{t_2},\hbar\omega_{t_3}$) are acquired by scanning and then applying  Fourier transforms with respect to the time delays $t_2$ and $t_3$, while holding $t_1$ constant as illustrated in Fig.~4a. Choosing $t_1=\unit[0]{fs}$, the zero-quantum spectrum of the MoSe$_2$ BL is presented in Fig.~4b. As a signature of population relaxation, we observe the main peak is distributed along the grey dashed line with $\hbar\omega_{t_2}\approx\unit[0]{meV}$. Examining a line cut through the peak of the exciton resonance along the $\hbar\omega_{t_2}$ direction (blue vertical dashed line), we can further extract the exciton population relaxation rate. Intriguingly, the profile in Fig.~4c could only be well-fitted with two Lorentzian functions with linewidths of 12.2 meV and 0.81 meV, respectively. Translating them to decay times, we obtained fast ($\tau_{fast}=\unit[54\pm2]{fs}$) and slow ($\tau_{slow}=\unit[810\pm10]{fs}$) components. These time scales are much faster than many previous reports on exciton population relaxation using pump/probe or time-resolved PL techniques \cite{2013_HShi&HLibai_ACSNano_T1_PumpProbe,2016_CRobert&XMarie_PRB_RadLifetimeMoSe2,2015_GWang&BUrbaszek_APL_TRPL_MoSe2ML}  because our experiments detect third-order coherent signals, enabling a quantitative comparison with microscopic calculations presented below. In contrast, incoherent spectroscopy techniques are often influenced by exciton-repopulation processes from defect-trapped states or conversion from indirect/dark excitons.

Our microscopic calculation begins by determining the Pauli blocking effect in each valley: the blocking is given by the temporal evolution of the overall carrier occupation in the $K$ valley $f = f^e + f^h$, which the third pulse is sensitive to. Electron and hole occupations are determined by the exciton states which have an electron/hole in the $K$ valley \cite{Katsch2018}: $f^e =  |P^{K-K}_0|^2 + \sum_{\mathbf{Q},i_h=K,\Gamma} N^{i_h K}_\mathbf{Q},$ and $f^h = |P^{K-K}_0|^2 + \sum_{\mathbf{Q},i_e=K,\Lambda,K',\Lambda'} N_\mathbf{Q}^{K i_e}$. We find contributions from the optically pumped coherent excitons $P^{K-K}_0$ as well as from incoherent excitons $N_\mathbf{Q}^{i_h i_e}$ formed through exciton-phonon scattering of coherent excitons \cite{Selig2018}. In our calculation of the temporal evolution of the coherent and incoherent excitons, we include exciton-photon, exciton-phonon and intervalley exchange interactions \cite{ 2019_MSelig&AKnorr_PRR_PhononValleyDarkExciton, Selig2018, selig2020suppression}. Our analysis (shown in Fig.~4d, details in SI) predicts a fast decay rate of \unit[12.1]{meV} (\unit[55]{fs}) originating primarily from the decay of coherent excitons, with additional contributions from the relaxation of $K-K$ excitons to momentum-indirect $K-\Lambda$ and $\Gamma-K$ states after the optical pump, and the further decay of $K-\Lambda$ excitons. The subsequent slow decay of \unit[0.85]{meV} (\unit[770]{fs}) is determined by the decay of the $\Gamma-K$ excitons to the $\Gamma-\Lambda$ exciton states. We present the calculated relaxation processes in the frequency and time domains as depicted in Fig.~4d and 4e, respectively. There is  excellent agreement between the experiments (Fig.~4c) and calculation (Fig.~4d). We re-plotted the calculated dynamics in the time domain for ease of visualization (Fig.~4e). In contrast, exciton population relaxation measured from a MoSe$_2$ ML (details included in SI) reveals a single component decay with a $\unit[475\pm8]{fs}$ relaxation time, an order of magnitude slower than the $\unit[54]{fs}$ BL component, emphasizing the distinct microscopic decay channels in the ML and BL.

\TODO{Early steady-state photoluminescence experiments on TMD bilayers identified additional exciton resonances attributed to electron and holes residing in different valleys.~\cite{kozawa2014photocarrier} Our study goes beyond previous works that suggested interlayer scattering processes should be considered in bilayers. We find that the emergence of additional low-energy valleys in MoSe$_2$ bilayers leads to rapid phonon-assisted inter-valley scattering, resulting in significantly faster intrinsic exciton dephasing and two components in the population relaxation dynamics.} Microscopic calculations allow us to attribute them to specific inter-valley scattering pathways involving $\Lambda$ valley in the conduction band and $\Gamma$ valley in the valence bands. Additional spectroscopy studies such as those based on time-resolved angle-resolved photoemission spectroscopy (Tr-APRES) measurements with momentum space resolution are needed \cite{madeo2020directly, 2020_S.Dong, 2021_Wallauer_ARPES} to directly visualize these inter-valley scattering processes. Understanding how low-energy valleys influence exciton quantum dynamics is critical to extending valleytronics in vdW heterostructures beyond the simplest case of ``spin-valley locking" found in TMDC MLs.

\section{Acknowledgement}

\bibliography{references}
\noindent[51] See Supplemental Material for sample images, theoretical treatment of the dephasing and relaxation mechanisms, and additional measurements, which includes Refs [52-65]\\
\noindent[52]
M. Selig, F. Katsch, R. Schmidt, S. M. de Vasconcellos, R. Bratschitsch, E. Malic, and A. Knorr, ArXiv:1908.10080 (2019).\\
\noindent[53]
N. S. Rytova, Proc. MSU, Phys., Astron. \textbf{3}, (1967).\\
\noindent[54]
S. Rudin, T. L. Reinecke, and B.Segall, Phys. Rev. B \textbf{42}, 17 (1990).\\
\noindent[55]
Trolle, Mads L., Pedersen, Thomas G., and Veniard, Valerie, Scientific Reports \textbf{7}, 39844 (2017).\\
\noindent[56]
Y. Li, A. Chernikov, X. Zhang, A. Rigosi, H. M. Hill, A. M. van der Zande, D. A. Chenet, E.-M. Shih, J. Hone, and T. F. Heinz, Phys. Rev. B \textbf{90}, 205422 (2014).\\
\noindent[57]
J. Lee, E. S. Koteles, and M. O. Vassell, Phys. Rev. B \textbf{33}, 5512 (1986).\\
\noindent[58]
A. Kormanyos, G. Burkard, M. Gmitra, J. Fabian, V. Zolyomi, N. D. Drummond, and V. Falko, 2D Materials \textbf{2}, 022001 (2015).\\
\noindent[59]
Z. Jin, X. Li, J. T. Mullen, and K. W. Kim, Phys. Rev. B \textbf{90}, 045422 (2014).\\
\noindent[60]
V. Srinivas, J. Hryniewicz, Y. J. Chen, and C. E. C. Wood, Phys. Rev. B \textbf{46}, 10193 (1992).\\
\noindent[61]
Z. Khatibi, M. Feierabend, M. Selig, S. Brem, C. Linderälv, P. Erhart, and E. Malic, 2D Materials \textbf{6}, 015015 (2018).\\
\noindent[62]
S. Helmrich, R. Schneider, A. W. Achtstein, A. Arora, B. Herzog, S. M. de Vasconcellos, M. Kolarczik, O. Schöps, R. Bratschitsch, U. Woggon, and N. Owschimikow, 2D Mater. \textbf{5}, 045007 (2018).\\
\noindent[63]
D. Erkensten, S. Brem, and E. Malic, ArXiv 2006.08392, (2020).\\
\noindent[64]
E. W. Martin, J. Horng, H. G. Ruth, E. Paik, M.-H. Wentzel, H. Deng, and S. T. Cundiff, ArXiv Preprint ArXiv:1810.09834 (2018).\\
\noindent[65]
T. Stroucken, A. Knorr, P. Thomas, and S. W. Koch, Phys. Rev. B \textbf{53}, 2026 (1996).\\

\end{document}